\documentclass[prb,twocolumn,aps]{revtex4}

\usepackage{graphicx}
\usepackage{latexsym}

\begin{document}

\title{High pressure diamond-like liquid carbon}

\author{Luca M. Ghiringhelli$^1$, Jan H. Los$^2$, Evert Jan Meijer$^1$,
A. Fasolino$^{1,2}$, and Daan Frenkel$^{1,3}$}

\affiliation{$^1$van 't Hoff Institute for Molecular Sciences, Universiteit van
Amsterdam, Nieuwe Achtergracht 166, 1018 WV Amsterdam, The
Netherlands\\
$^2$Theoretical Physics, NSRIM, University of Nijmegen,
Toernooiveld, 6525 ED Nijmegen, The Netherlands\\
$^3$FOM Institute for Atomic and Molecular Physics,
Kruislaan 407, 1098 SJ, Amsterdam, The Netherlands}

\date{\today}

\begin{abstract}
  We report density-functional based molecular dynamics simulations,
  that show that, with increasing pressure, liquid carbon undergoes a
  gradual transformation from a liquid with local three-fold
  coordination to a 'diamond-like' liquid.  We demonstrate that this
  unusual structural change is well reproduced by an empirical bond
  order potential with  isotropic long range interactions, supplemented by
  torsional terms. In contrast, state-of-the-art short-range
  bond-order potentials do not reproduce this diamond structure.  This
  suggests that a correct description of
  long-range interactions is crucial for a unified
  description of the solid and liquid phases of carbon.
\end{abstract}

\maketitle
\begin{figure}[b]
\includegraphics[width=8.5cm]{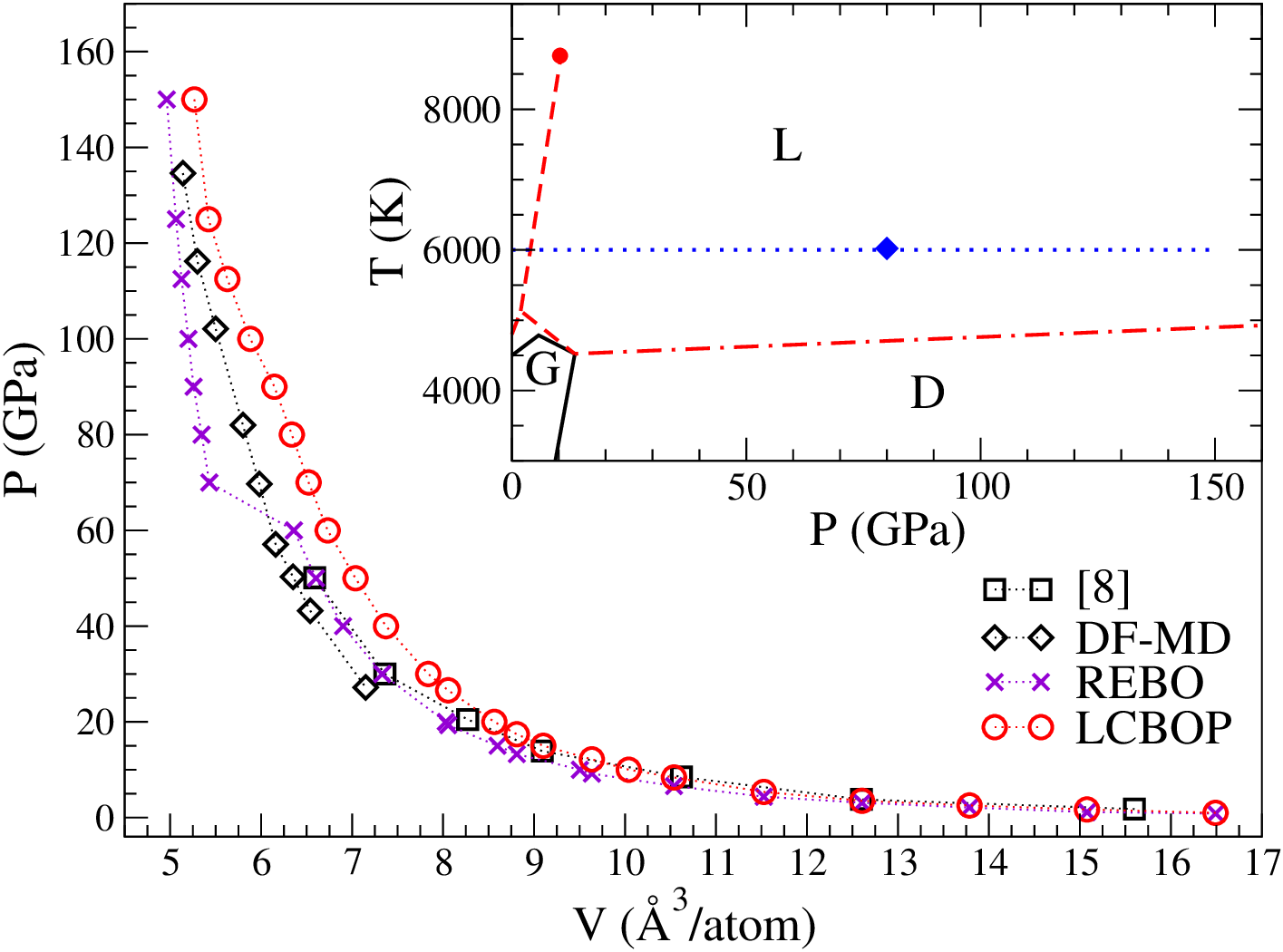}
\vspace*{-0mm}
\caption{(Color online) Equation of state of liquid carbon at T=6000K.
The inset shows the $P-T$ phase diagram, with diamond (D), graphite
(G) and liquid (L) domains: the dotted curve is the sampled
isotherm; the solid curves are experimental data: the
diamond-graphite coexistence curve from Bundy~\cite{Bundy}, the
graphite melting curve from Togaya~\cite{Togaya}. The dot-dashed line is
the estimated diamond melting line from
Glosli and Ree~\cite{GlosliReeMelting}, obtained with Brenner potential; the
dashed curves are the liquid-liquid coexistence curve ending in a
critical point and its graphite melting curve from Glosli and
Ree~\cite{Glosli}.
The solid $\diamond$ in the inset indicates the switching in coordination shown in
Fig.~\ref{Coord}.} \label{EqState}
\end{figure}

When a simple fluid, such as argon, is cooled below the critical
temperature, it can form a liquid phase. However, more complex
fluids, such as molten phosphorus, can form two distinct liquid
phases that are separated by a first-order  liquid-liquid phase
transition (LLPT)~\cite{Katayama}.
There is increasing evidence that a variety of network-forming
liquids can undergo an LLPT (see e.g. Franzese {\em et al.}~\cite{Stanley}).
In particular, it has long been suspected that elemental carbon
may exhibit an LLPT~\cite{Ferraz,vanThiel}. As direct experiments
are difficult in the relevant temperature regime (4000 to 6000 K),
much of the evidence for such a transition comes from simulations.
Classical simulations based on the Brenner bond-order potential
with torsional terms~\cite{Brenner,BrennerTors} predicted a
coexistence line between a 2-fold and a four-fold coordinated
liquid starting from the melting line of graphite end ending in a
critical point (at $\sim 8800$~K)~\cite{Glosli} as shown in
Fig.~\ref{EqState}. Recent first-principle
investigations~\cite{Wu} have ruled out the occurrence of such a
transition in the  range of  specific volumes 6.59 -
15.6~\AA$^3$/atom at 6000 K. More importantly, these ab-initio
results found a majority of three-fold coordination over the
studied range. Wu {\em et al.}~\cite{Wu} pointed out  that the presence
of $\pi$-bonds in the liquid phase makes it difficult to represent
the torsion potential adequately. They concluded that empirical
potentials are unable to capture these subtle electronic
effects.

In this Rapid Communication, we reexamine these issues  by extending the range
of studied densities   while comparing density functional theory
based molecular dynamics (DF-MD)  simulations to  Monte Carlo (MC)
results based on two state-of-the-art empirical potentials. The
first is the recently proposed long-range carbon bond-order
potential (LCBOP~\cite{Jan}) extended with torsional interactions;
the second is the reactive empirical bond-order
(REBO\cite{BrennerREBO}) potential which improves the earlier
version due to Brenner~\cite{Brenner}. LCBOP introduces long range
terms to account for interplanar binding of graphite which is not
described by the REBO potential.

On the basis of DF-MD we find that, whilst no true LLPT occurs
even for higher densities than in Wu {\em et al.}~\cite{Wu}, a well defined
switching of the dominant coordination from three- to four is found
at  $\sim$ 5.8 ~\AA$^3$/atom, just  beyond the range studied  in
Wu {\em et al.}~\cite{Wu}. This 'diamond-like' four-fold coordinated liquid
is  highly structured, with a strongly anisotropic angular
distribution of the first neighbors. Interestingly, the empirical
potential LCBOP gives an impressive agreement with the DF-MD
results for the structural properties of this high density liquid
whereas the short-ranged REBO potential~\cite{BrennerREBO} yields
a graphite-like liquid in this range. It should be noted that both
the low-density three-fold coordinated and high density
'diamond-like' liquid are very unusual. In fact most covalent
semiconductors become highly coordinated in the liquid
phase~\cite{GeSiliquid}. Highly coordinated phases have been
predicted by Grumbach~\cite{CHiP} also for carbon albeit at
pressures of TPa.

We performed constant volume DF-MD simulations using the
Car-Parrinello method~\cite{CP} as implemented in CPMD
package~\cite{CPMD}.  The system consisted of 128 atoms in a cubic box
with periodic boundary conditions at 9 densities and a temperature
$T=6000$~K, imposed by means of a Nos\'e-Hoover~\cite{N-H} thermostat.
We used the Becke~\cite{Becke} exchange and
Perdew~\cite{Perdew} correlation gradient corrected functional (BP)
with a plane wave basis set cut off at 35 Ry and sampled the Brillouin
zone only in the gamma point.  BP gives a correct description of bulk
diamond.  Each state point was studied for 5 ps, starting from a
sample equilibrated via LCBOP; only minor structural changes occurred
in the first tenths of ps.  Since liquid carbon is metallic, we
imposed a thermostat for the electronic degrees of freedom in order to
ensure a proper implementation of the Car-Parrinello
scheme~\cite{AdiabCP}.  We performed the MC simulations of 128 particles in a
cubic box with periodic boundary conditions with the LCBOP and the
REBO potential. We sampled at 6000 K the constant volume ensemble for
specific volumes larger than 8 \AA$^3$/atom and the constant pressure ensemble
for smaller specific volumes where the increase of pressure is steeper, with an overlap
between the two regions to check for consistency.  MC simulations at
selected volumes with 512 and 4096 particles using the LCBOP showed
negligible differences with the 128 particle results for the local
structure and equation of state.

In order to reproduce the ab-initio results, we needed to refine
the LCBOP. The detailed procedure  will be described
elsewhere~\cite{us}. Here we give a brief summary of the changes.
Following the strategy of the REBO potential~\cite{BrennerREBO} we
have introduced a correction to the angular function in order to
stabilize small clusters in shapes other than chain-like. It is in
fact known (see e.g. Raghavachari and Binkley~\cite{Rhombus}) that,
while odd numbered clusters prefer to arrange in a chain, even numbered
ones can also have metastable ring or cage structures. We focused
attention on a planar $C_4$ cluster with $C_{2v}$ symmetry (a rhombus with
two $\sim 60^{\circ}$ angles\cite{Rhombus}) and a cubic $C_8$\cite{Cube}.
It turned out that the angular dependence of the
bond order had to be weakened for coordination lower than three.
More important is the addition of torsional interactions, in such
a way as to describe the conjugation dependence of the torsional
energy barrier in agreement with the ab-initio calculations for
double and conjugated bonds of Wu {\em et al.}~\cite{Wu}. This dependence
 is not captured by the REBO potential.
There, the barrier as a function of the dihedral angle
$\omega_{ijkl}$ is always described by $sin^2(\omega_{ijkl})$ and
 differs only by a scale factor between these two bonding situations.
In the original LCBOP, a variable $N_{ij}^{conj}$ was already defined,
assuming values 1 or 0 respectively in these two extreme cases,
to account for the presence or absence of a $\pi$ bond.
We fit the results of Wu {\em et al.}~\cite{Wu}  for $N_{ij}^{conj}=0,1$
by two polynomials in $cos^2(\omega_{ijkl})$
and interpolate between them by means of a function decaying rather quickly
away from $N_{ij}^{conj}=1$.
The torsional energy is activated only for a bond between three-fold
coordinated atoms, as for the REBO potential.
The smaller torsional energy of a bond between four-fold coordinated atoms
is already taken into account
by the long range interactions between the third neighbours.

\begin{figure}[t]
\vspace*{5mm}
\includegraphics[width=8cm]{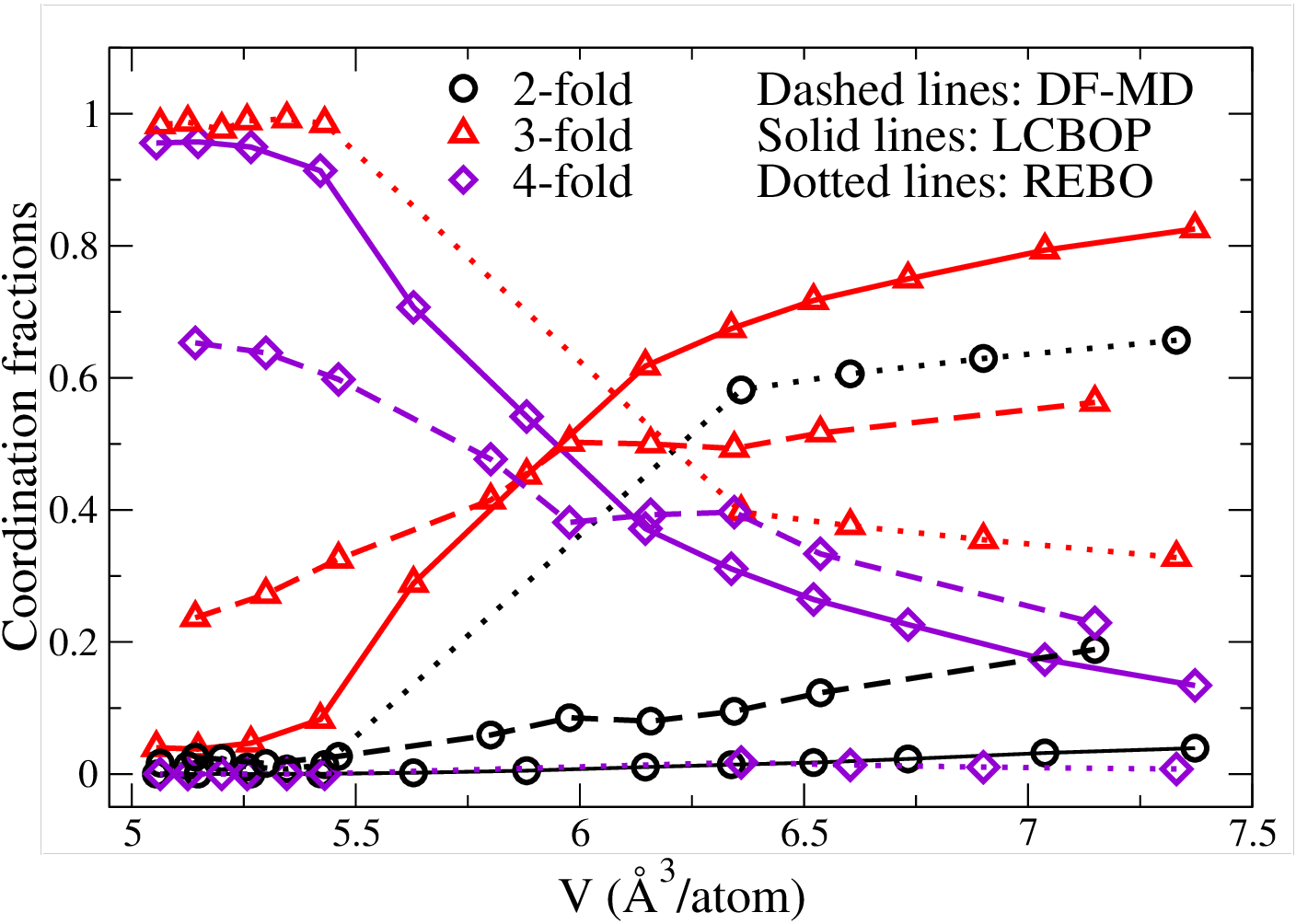}
\caption{(Color online) Coordination fractions versus specific volume
of liquid carbon.
Note that at  $V\sim 6$ \AA$^3$/atom DF-MD and LCBOP predict a
switch from three- to four-fold coordination whereas REBO evolves
from a 2-fold to a graphite like liquid.}
\label{Coord}
\end{figure}

In Fig.~\ref{EqState} we compare the pressure-volume isotherms of
our DF-MD simulations with our MC results based on the empirical
REBO potential and LCBOP, and with the DF-MD data of
Wu {\em et al.}~\cite{Wu}, all at a temperature of 6000 K. In view of the
relatively low cut-off  (35 Ry), we had to correct the pressures
for the spurious contribution due to  Pulay forces~\cite{Pulay}.
In the density range where we can compare with the results of
Wu {\em et al.}~\cite{Wu}, the  pressures that we compute are some
15\% lower than those reported by Wu {\em et al.} This relatively
small difference is probably due to the different choice of
density functionals. We have checked that our samples were indeed
liquid. Over the whole isotherm we have observed diffusive
behavior in both the MC-LCBOP and the DF-MD simulations, the
latter indicating a self-diffusion coefficient at least of order
$10^{-5} $cm$^2$/s.

\begin{figure}[h]
\vspace*{5mm}
\includegraphics[width=7cm]{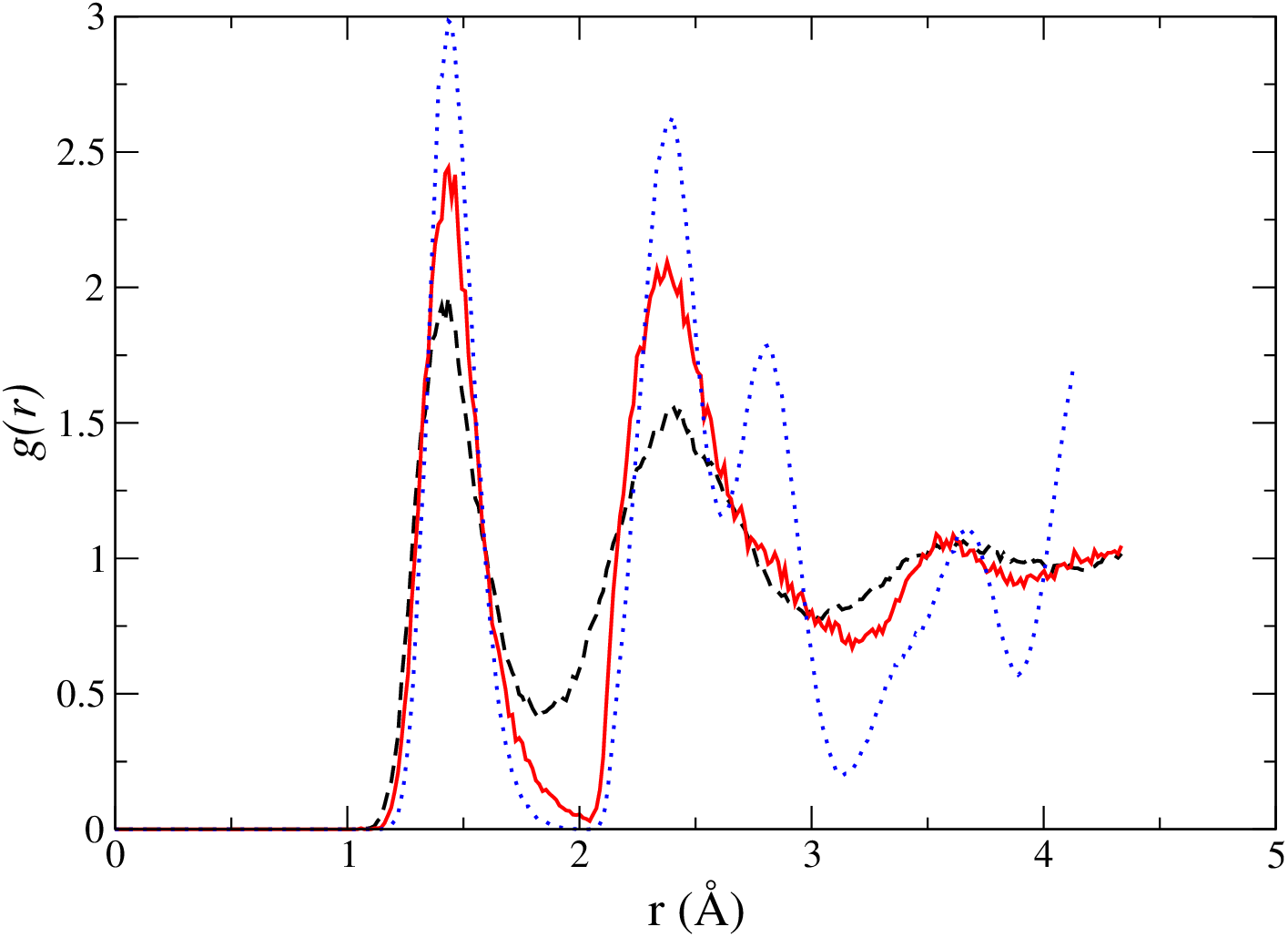}
\caption{(Color online) Pair correlation functions at 5000 K. Diamond
(dotted curve) is at 150 GPa; the LCBOP liquid (solid curve) is at 150 GPa
and 5.14~\AA$^3$/atom; DF-MD liquid (dashed curve) is at
5.14~\AA$^3$/atom.} \label{RDF}
\end{figure}

Up to 60 GPa both empirical potentials seem to be in fair
agreement with ab-initio data. However, the coordination fractions
shown in Fig.~\ref{Coord} indicate that the REBO potential fails
to reproduce the correct liquid structure. It predicts, in the range
between 6 and 7.5 \AA$^3$/atom, mainly
2-fold coordination
whereas the present DF-MD results
(and those of  Wu {\em et al.}~\cite{Wu}), as well as the LCBOP,
show dominant three-fold coordination and low 2-fold coordination.

At higher pressures, DF-MD predicts a marked switching of dominant
coordination from three to four around 5.8 \AA$^3$/atom.
Judging from the isotherm of Fig.~\ref{EqState}, the transition
seems to be continuous with no sign of a van der Waals loop. These
results are consistent with the tight binding MD simulations of
Morris {\em et al.}~\cite{Morris}. In contrast, between 5.6 and
6.0~\AA$^3$/atom, where the switch of dominant coordination takes
place, the MC results based on LCBOP display large fluctuations in
density at the imposed pressure of 100 GPa, resulting in a slight bending
of the isotherm of Fig.\ref{EqState}.
Again, qualitatively the coordination fractions obtained by LCBOP
agree with the DF-MD results even though with a more pronounced
switch to coordination four. Curiously, the REBO potential
predicts a first order phase transition in the same density range
but to a completely different structure, a three-fold
graphite-like liquid with well ordered sliding sheets which
eventually get stuck upon further increasing of the pressure.
The REBO potential never gives rise to four-fold coordination.

If the three- to four-fold coordination change extrapolates to
lower temperatures, it might provide an explanation for the sudden
change of the slope of the graphite melting line~\cite{Togaya}.
The origin of the latter is still an open question, although it
has been suggested that it is associated with a first order phase
transition~\cite{vanThiel,Togaya}. Our results provide no evidence
of a first-order transition but rather indicate a pronounced but
continuous change of dominant coordination.

\begin{figure}[h]
\vspace*{5mm}
\includegraphics[width=7cm]{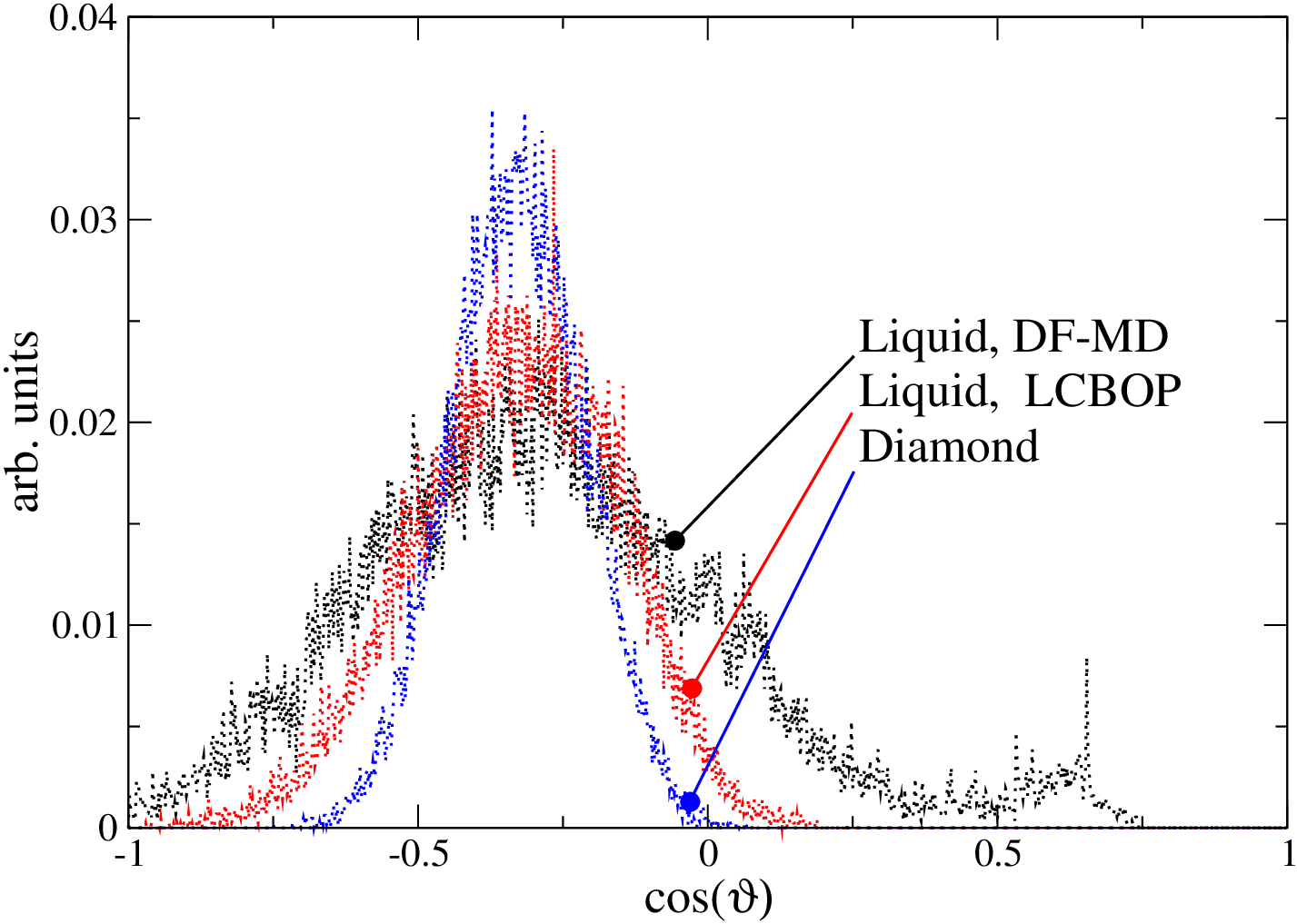}
\caption{(Color online) Angular correlation functions of first neighbors
with state parameters as in Fig.~\ref{RDF}.}
\label{ang1stshell}
\end{figure}

\begin{figure}
\includegraphics[width=7cm]{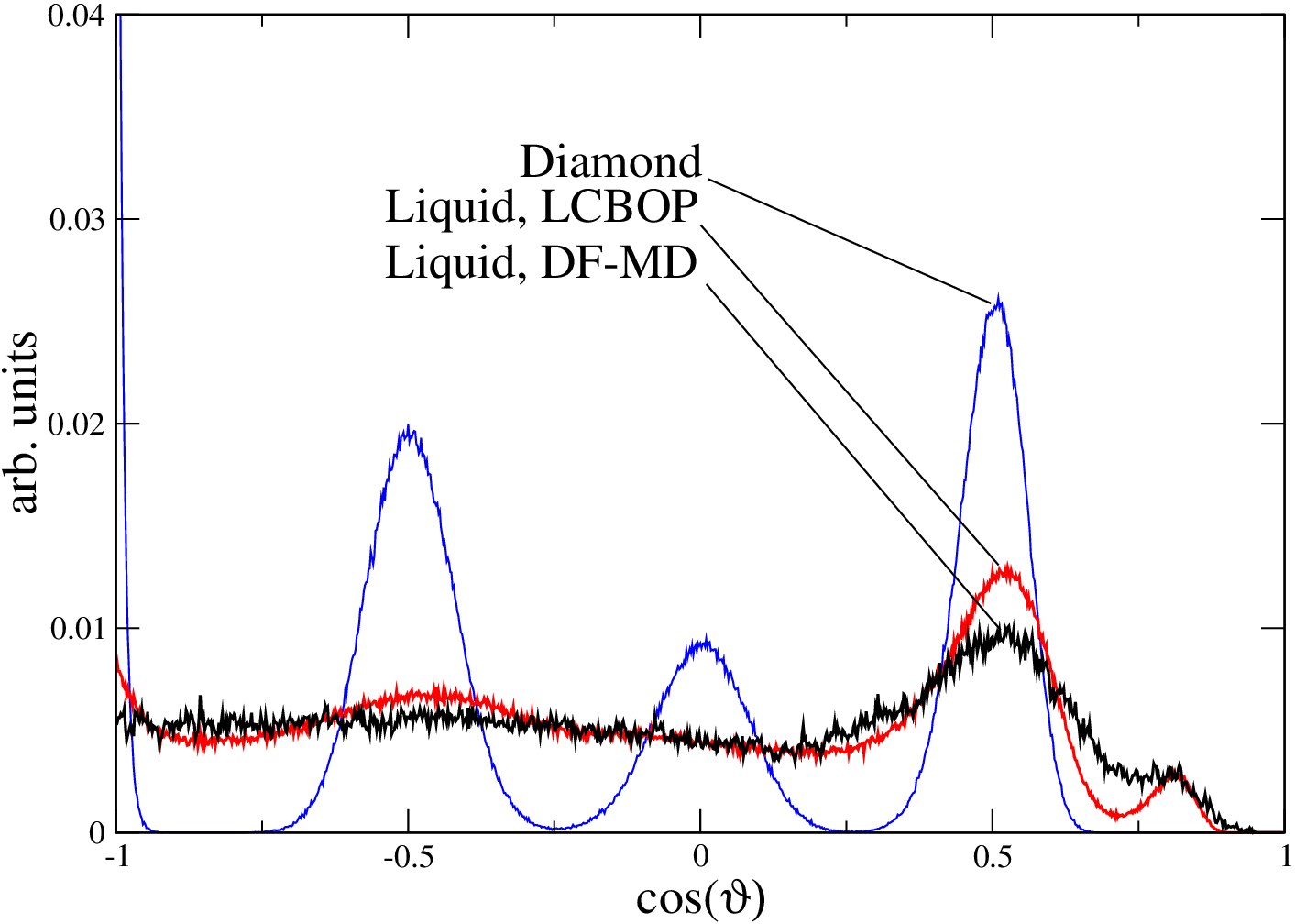}
\caption{(Color online) Angular correlation functions for second neighbors
(see text) with state parameters as in Fig.~\ref{RDF}
The peaks for the diamond are spread around the theoretical
values for an fcc lattice of -1, -0.5, 0, 0.5, with weights
6/66, 24/66, 12/66, 24/66.}
\label{ang2ndshell}
\end{figure}

Both the DF-MD and LCBOP simulations of the high density four-fold
coordinated liquid show a strong diamond like positional and
orientational order, as shown in Figs.~\ref{RDF}, \ref{ang1stshell},
and \ref{ang2ndshell}.
In Fig.~\ref{RDF}, we compare the pair correlations $g(r)$ of this
high density liquid with that of a (meta)stable bulk diamond near the
estimated coexistence point, at 5000K and 150 GPa (also the LCBOP
and DF-MD liquid samples were equilibrated at that temperature).
One can see that up to the second neighbor shell the liquid has a structure
almost as pronounced as diamond. The $g(r)$'s obtained by LCBOP globally
agree fairly well with the ones obtained by DF-MD~\cite{us},
except around  2 \AA, where the LCBOP minimum is too deep.
In Fig.~\ref{ang1stshell} we present the calculated angular
correlation $g^{(3)}(\theta)$ for first neighbors, i.e. those
atoms that fall within the short range cut-off of the LCBOP.
Again, the first shell of neighbors in the liquid has a strong
tetrahedral ordering, comparable to bulk diamond.
To test how far the local diamond structure persists in the
liquid,  we define the angular correlation function for second
neighbors (i.e. all those particles that are first neighbors of
the first neighbors, excluding the central atom and the
first-neighbor shell). Fig.~\ref{ang2ndshell} shows that, in the
second-neighbor shell, the diamond structure is completely lost.
Yet, the angular distribution is not structureless:  we find a
peak around $60^{\circ}$ and a shoulder at $\sim 35^{\circ}$; in a
diamond lattice, the latter feature can be attributed to
cross-correlations between the first and the second neighbor
shells.

It is rather surprising that the LCBOP potential reproduces the
transformation to a predominantly four-fold coordinated liquid,
while the REBO potential does not. Apparently, the
isotropic long-ranged interactions play a
crucial role. To demonstrate this, we verified that he
short-ranged CBOP reproduces the behavior of the REBO
potential~\cite{Jan}. This behavior is rather puzzling as
long-ranged interactions are expected to play a negligible role at
these high densities (see, e.g., Glosli and Ree~\cite{GlosliReeMelting}).
Moreover, long-ranged interactions were introduced in
Los and Fasolino~\cite{Jan} to describe three-fold coordinated
graphitic phases, and no attempt was made to make the long range
interactions dependent on the local environment.
Torsional interactions appear to important, since, without them,
the calculated pressures would be too high for high densities and
too low at low densities.

We conjecture that the combination of torsional interactions, and long
range forces is required to give the best description of the liquid.
It would be interesting to check this conjecture by studying the
liquid with other empirical potentials, in particular with Marks'
environment-dependent interaction potential (EDIP)~\cite{Marks}. The
study of the liquid with this potential presented in
Marks~\cite{Marks2} is limited to volumes of 6.2~\AA$^3$/atom where it
gives coordination fractions comparable to those obtained here by DF
and by LCBOP and in Wu {\em et al.}~\cite{Wu}.

In summary, on the basis of DF-MD simulations, we predict that
liquid carbon should exhibit a continuous transformation from a
three-fold to a mostly four-fold coordinated liquid at high
pressure. This liquid  has a strong local angular ordering,
reminiscent of the diamond structure. We have compared this
finding with the results of MC simulations based on two empirical
bond-order potentials with torsional terms, the short-range REBO
potential and the long-range LCBOP. We find that the latter
reproduces quite well the structure of both low-density and
high-density liquid. The REBO simulations instead display a marked
first order phase transition between a mostly 2-fold and a
mostly three-fold graphite fluid in this range. It is tempting to
speculate that the existence of a four-fold coordinate liquid at high
densities will greatly facilitate the nucleation of diamonds from
dense liquid carbon.

This work is part of the research program of the 'Stichting voor
Fundamenteel Onderzoek der Materie (FOM)', which is financially
supported by the 'Nederlandse Organisatie voor Wetenschappelijk
Onderzoek (NWO)'.  E.J.M. acknowledges the Royal Netherlands Academy
of Art and Sciences for financial support.  J.L. and A.F. acknowledge
NWO project 015.000.031 for financial support.  We acknowledge support
from the Stichting Nationale Computerfaciliteiten (NCF) and the
Nederlandse Organisatie voor Wetenschappelijk Onderzoek (NWO) for the
use of supercomputer facilities.


\clearpage
\newpage


\begin{thebibliography}{99}
\bibitem{Katayama} Y. Katayama, T. Mizutani, W. Utsumi, O. Shimomura,
M. Yamakata, K. Funakoshi, Nature {\bf 403}, 170 (2000).
\bibitem{Stanley} G. Franzese, G. Malescio,  A.Skibinsky, S.V. Buldyrev,
H.E. Stanley, Nature {\bf 409}, 692 (2001).
\bibitem{Ferraz} A. Ferraz, N.H. March, Phys. Chem. Liq. {\bf 8}, 289 (1979).
\bibitem{vanThiel} M. van Thiel, F. H. Ree Phys. Rev. B {\bf 48}, 3591 (1993).
\bibitem{Brenner} D.~W. Brenner, Phys. Rev. B {\bf 42}, 9458 (1990).
\bibitem{BrennerTors} D.~W. Brenner, J.~A. Harrison, C.~T.White,
R.~J. Colton, Thin Solids Films {\bf 206}, 220 (1991).
\bibitem{Glosli} J.~N. Glosli and F.~H. Ree, Phys. Rev. Lett. {\bf 79},
2474 (1997).
\bibitem{Wu} C.~J. Wu, J.~N.Glosli, G. Galli, F.~H. Ree,
Phys. Rev. Lett. {\bf 89}, 135701 (2002).
\bibitem{Jan} J.~H. Los and A. Fasolino, Phys. Rev. B {\bf 68}, 024107 (2003).
\bibitem{BrennerREBO} D.~W.Brenner, O.~A. Shenderova, and J.~A. Harrison,
et al, J. Phys. - Condens. Mat. {\bf 14}, 783 (2002).
\bibitem{Bundy}  F.~P. Bundy, J. Chem. Phys {\bf 38}, 618 (1963).
\bibitem{Togaya} M. Togaya, Phys. Rev. Lett. {\bf 79}, 2474 (1997).
\bibitem{GlosliReeMelting} J.~N. Glosli and F.~H. Ree, J. Chem. Phys.
{\bf 110}, 441 (1999).
\bibitem{GeSiliquid} M.~P. Tosi, J. Phys-Cond. Matter {\bf 6}, A13-A28 (1994).
\bibitem{CHiP} M.~P. Grumbach and R.~M. Martin, Phys Rev. B {\bf 54},
15730 (1996).
\bibitem{CP} R. Car and M. Parrinello, Phys. Rev. Lett. {\bf 55}, 2471 (1985).
\bibitem{CPMD} CPMD, J. Hutter, A. Alavi, T. Deutsch, M. Bernasconi,
St. Goedecker, D. Marx, M. Tuckerman, M. Parrinello, MPI f\"ur
Festk\"orperforschung and IBM Zurich Research Laboratory 1995-1999.
\bibitem{N-H} W.~G.Hoover, Phys. Rev. A {\bf 31}, 1695 (1985).
\bibitem{Becke} A.~D. Becke, Phys. Rev. A {\bf 38}, 3098 (1988).
\bibitem{Perdew} J.~P. Perdew, Phys. Rev. B {\bf 33}, 8822 (1986),
Erratum Phys. Rev. B {\bf 34}, 7406 (1986).
\bibitem{AdiabCP} P.~E. Bl\"ochl and M. Parrinello, Phys. Rev. B
{\bf 45}, 9413 (1992).
\bibitem{us} J.~H. Los {\em et al.}, in preparation.
\bibitem{Rhombus} K. Raghavachari and J.~S. Binkley, J. Chem. Phys. {\bf 87},
2191 (1987).
\bibitem{Cube} K. Kobayashi, N. Kutita, H. Kumahora, K. Tago, Phys Rev. B
{\bf 45}, 11299 (1992).
\bibitem{Pulay} P.~G. Dacosta, O.~H. Nielsen, and K. Kunc,
J. Phys. C: Solid State Phys. {\bf 19}, 3163 (1986).
\bibitem{Morris} J.~C. Morris, C.~Z. Wang, and K.~M. Ho, Phys. Rev. B
{\bf 52}, 4138 (1995).
\bibitem{Marks}  N.~A. Marks, Phys. Rev. B {\bf 63} , 035401 (2001).
The EDIP potential takes in account interactions within a radius of 3.2 \AA.
\bibitem{Marks2} N. Marks, J. Phys.: Condens. Matter {\bf 14}, 2901 (2002).
\end{thebibliography}
\end{document}